\documentclass[journal=jctcce,manuscript=article,layout=traditional]{achemso}

\usepackage[T1]{fontenc} %
\usepackage[version=3]{mhchem} %
\usepackage{xcolor}
\usepackage{physics}
\usepackage{overpic}
\usepackage{soul} 

\newcommand{\wn}{\ifmmode\;\text{cm}^{-1}\else$\text{cm}^{-1}$\fi}
\newcommand{\mc}[1]{\ensuremath{\mathcal{#1}}}
\newcommand{\mat}[1]{\ensuremath{\mathbf{#1}}}

\newcommand{\reffig}[1]{Fig.~\ref{#1}}
\newcommand{\fig}[1]{Fig.~\ref{#1}}
\newcommand{\refeq}[1]{Eq.~\ref{#1}}

\newcommand{\unigro}{Zernike Institute for Advanced Materials, Faculty of Science and Engineering, University of Groningen, Nijenborgh 4, 9747AG Groningen The Netherlands.}

\newcommand{\upenn}{Chemistry, University of Pennsylvania, 231 S. 34 Street, Cret Wing 141D, Philadelphia, Pennsylvania 19104-6243, United States}

\newcommand{\uokla}{Department of Chemistry and Biochemistry, University of Oklahoma, Norman, Oklahoma 73019, United States}

\newcommand{\casai}{AI for Science Institute, Beijing 100080, China}

\author{D. Vale Cofer-Shabica}
\email{valecs@sas.upenn.edu}
\affiliation[upenn] {\upenn}
\altaffiliation{Contributed equally to this work}
\author{Maximilian F.S.J. Menger}
\affiliation[unigro] {\unigro}
\altaffiliation{Contributed equally to this work}
\author{Qi Ou}
\affiliation[casai] {\casai}
\author{Yihan Shao}
\affiliation[uokla] {\uokla}
\author{Joseph E. Subotnik}
\affiliation[upenn] {\upenn}
\author{Shirin Faraji}
\email{s.s.faraji@rug.nl}
\affiliation[unigro] {\unigro}

\title[INAQS]{{INAQS}, a generic interface for non-adiabatic {QM/MM} dynamics: Design, implementation, and validation for GROMACS/Q-CHEM simulations}
\abbreviations{QMMM, SH}
\keywords{QMMM, Surface Hopping}

\begin{document}

\begin{tocentry}
\centering
\includegraphics[width=0.62\textwidth]{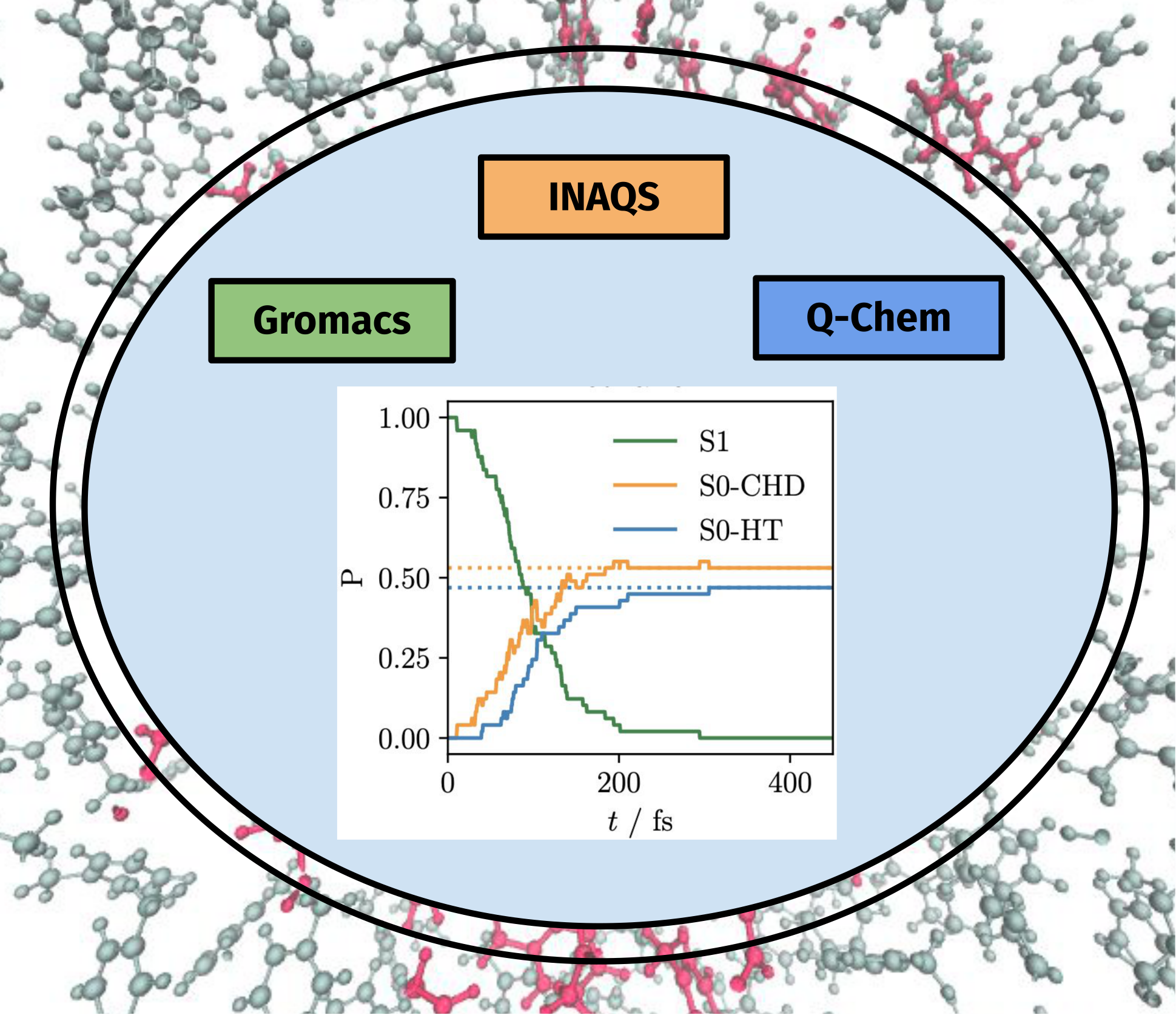}
\end{tocentry}

\begin{abstract}
The accurate description of large molecular systems in complex environments remains an ongoing challenge for the field of computational chemistry. This problem is even more pronounced for photo-induced processes, as multiple excited electronic states and their corresponding non-adiabatic couplings must be taken into account. Multiscale approaches such as hybrid quantum mechanics/molecular mechanics (QM/MM) offer a balanced compromise between accuracy and computational burden. 
Here, we introduce an open-source software package (INAQS) for non-adiabatic QM/MM simulations that bridges the sampling capabilities of the GROMACS MD package and the excited-state infrastructure of the Q-CHEM electronic structure software. The interface is simple and can be adapted easily to other MD codes. The code supports a variety of
different trajectory based molecular dynamics, ranging from Born-Oppenheimer to surface hopping dynamics.
To illustrate the power of this combination, we simulate electronic absorption spectra, free energy surfaces along a reaction coordinate, and the excited state dynamics of  1,3-cyclohexadiene in solution.
\end{abstract}

\section{Introduction}
The accurate description of large supramolecular and solvated systems presents a considerable challenge in the field of computational chemistry.
Given the inherent electronic structure difficulties of modeling bond-breaking and charge reorganization  in vacuum, it follows that modeling such processes in the presence of a complex environment can be quite difficult.
In the case of  extended systems, a complete quantum description is still not feasible today; even in light of the algorithmic and computational advances promised by new computational schemes\cite{xtb,neese2016:DLPNOgrad,akimov_large-scale_2015,groenhof2015:qmmmPMF} and GPU acceleration.\cite{terachem,brianQC,g16b01}
Luckily, for many chemical processes of interest, quantum effects  occur within a spatially localized region; and for such systems---provided the dynamics occur along the  electronic \emph{ground state}---there does exist today an enormous quantum mechanics/molecular mechanics (QM/MM) computational infrastructure.

At the frontier of modern QM/MM software is the study of non-equilibrium processes, especially non-adiabatic \textit{excited-state} molecular dynamics (e.g., photoreactions in light-harvesting complexes\cite{Curutchet2017,martinez2021:gfp,zimmer2002:GFPreview,jailaubekov2013hot}) in the presence of a complex environment\cite{barbatti2018:NAQMMMreview}.
For excited-state problems, the role of the environment is even more important than for the ground state. After all,  photochemical processes can release a great deal of energy that must be dissipated or marshaled by the environment.
Unfortunately, however, when simulating nonadiabatic dynamics with a strongly interacting environment, new theoretical issues do arise. To see this point, consider a typical system with heavy nuclei and light electrons. For such a system,  one can safely assume the electrons move with nuclei frozen or quasi-frozen, but the perennial question in nonadiabatic dynamics is: how does one move the {\em nuclei} if one does not know the correct electronic state to move along?  Within the context of QM/MM approaches, this question must then be superimposed with yet another question: how does one best move the nuclei of the system in the presence of solvent nuclei? In short, the presence of QM/MM  interactions introduces more time scales and potentially more questions to a nonadiabatic dynamics problem.

The answers to these questions inevitably depend on what level of theory one considers and what approximations one is willing to make. As far as treating the solvent environment, there are 
basically two approaches, implicit and explicit:
\begin{itemize}
\item Implicit models, such as the polarizable continuum model (PCM),\cite{Miertus1982,Mennucci2007,Mennucci:2012ct} the conductor like screening model (COSMO)\cite{Klamt_JCS_Cosmo,Pye1999TCA}, or SMx models,\cite{marenich_universal_2009,marenich_generalized_2013} describe the environment in terms of a continuous medium. 
These approaches are very computationally efficient and best suited for situations where the environment is homogeneous since a detailed molecular structure of the solvent is not given. To date, however, theoretical challenges remain as far as safely merging nonadiabatic dynamics with implicit solvation models. For instance, most semiclassical approaches for nonadiabatic dynamics rely on a time scale separation between slow nuclei and fast electrons; but for PCM models, one assumes the solvent response is also very fast, which can complicate the necessary dynamical question. 

\item Explicit QM/MM solvation approaches directly include individual solvent molecules in the simulation, often through classical force fields (FFs) and molecular mechanics (MM).
These approaches are more computationally expensive than the implicit models because of their sampling requirements, but can better treat heterogeneous environments and situations where the coupling between the active site and environment depends on the relative orientation or position of the active site and the immediate environment. Moreover, because one treats system nuclei and solvent nuclei at the same level of theory, there are fewer theoretical challenges as far as implementing nonadiabatic semiclassical dynamics algorithms. 
\end{itemize}

Now, the tradeoff for substituting explicit for implicit QM/MM models is that one trades in theoretical problems for practical questions:
How can one  accurately treat the interaction between QM and MM subsystems\cite{Warshel1976, Maseras1995, Svensson1996, Dapprich1999, Morokuma2006, Lin2006, Groenhof2013, Loco:2016bz, Curutchet2017}?
How should one treat the boundary, especially when a covalent bond bridges the interface\cite{Lin2005, Zhang2007, Boulanger:2012gx}?
A successful QM/MM software package must combine complicated electronic structure and molecular mechanics calculations, and problems often arise when there is tight coupling between the different components; moreover, a useful software package must be well structured, adaptable, and highly efficient given the large spatial scales of the systems of interest and the need to propagate simulations lasting long times.

With these general concerns in  mind, the existing QM/MM implementations can be roughly divided into three different broad categories:

\begin{enumerate}
\item \label{scheme:qm} \textbf{QM Driver} A straight-forward approach to QM/MM simulations is to implement all the necessary code within an existing electronic structure software package (e.g. Q-Chem~\cite{qchem}, Gaussian~\cite{g16b01}).
Here, the main driver is the QM code, which requires an internal implementation of MM forcefields and any dynamics desired.
In these cases the focus is typically on the description of the QM subsystem and optimizations or scans can use specialised approaches like the microiteration scheme\cite{kastner_exploiting_2007}.
Molecular dynamics is generally not the main focus and the statistical capabilities are often less comprehensive than in MD codes.

\item \label{scheme:md}\textbf{MD Driver} Another approach is to implement the necessary code as part of an existing MD code like GROMACS\cite{gromacs45}, NAMD,\cite{melo_namd_2018} AMBER\cite{Walker2008, Goetz2014} or CHARMM \cite{Charmm}.
Here, the QM contributions are typically provided by an interface to external QM codes. This approach is built around the molecular dynamics capabilities of the MD codes and enables advanced sampling techniques for various ensembles and QM/MM dynamics.
Features for static calculation, like optimizations or transition state searches are often limited, though in the condensed phase they are of questionable utility.

\item \label{scheme:standalone} \textbf{Standalone} A third approach is a fully independent implementation: constructing a new program that collates the outputs of existing QM and MM codes to build up a QM/MM calculation.
This is immensely flexible because  the developer has full control over the capabilities and can interface with whatever QM and MM codes she or he chooses.
However flexibility comes at a price: the developer must re-implement any desired features for dynamics, optimization, or statistics.
Examples of such hybrid software packages include ChemShell\cite{QUASI, ChemShell, PyChemShell}, CobraMM\cite{CobraMM2007, CobraMM2018}, QMMM\cite{QMMM2018}, SHARC\cite{Richter2011, Mai2018}, and Newton-X\cite{Newtonx, NewtonxQMMM}.
\end{enumerate}

To date, for ground state properties, a large variety of implementations exist employing all three of the above approaches.
For excited-state properties, especially when nonadiabatic effects must be taken into account, most implementations follow either the QM driver (\ref{scheme:qm}) or standalone (\ref{scheme:standalone}) schemes.
In particular, for nonadiabatic dynamics including solvent effects, the electronic properties (e.g. energies, gradients, nonadiabatic couplings) are typically computed via the standalone approach\cite{NewtonxQMMM,Mai2018,Thiel2014,Granucci2007,Thiel2011,Davide2021,Menger2018,Persico2014}. 
To the authors knowledge, and with the exception of some in-house codes\cite{groenhof2004:excitedQMMM,morozov2016photobiology}, integrating the nonadiabatic effects into an existing MD code (\ref{scheme:md}) is  rarely employed.\cite{Ivano2011}
But, this approach has several distinct advantages: reduced data transfer and efficient schemes for storing and analyzing long trajectories; the widespread availability of established FFs for solvents of all kinds; the use of advanced sampling and metadynamic schemes.

With all of this background in mind, here we present INAQS (Interface for Non-Adiabatic Quantum mechanics in Solution), a new interface for non-adiabatic QM/MM dynamics following approach (\ref{scheme:md}).
INAQS links the MD code Gromacs to the electronic structure software Q-Chem and enables ground- and excited-state ab-initio MD (AIMD), nonadiabatic surface hopping and Ehrenfest dynamics, and enhanced sampling within the mechanical or electrostatic embedding QM/MM framework.
Below, we present the software implemented, a few applications to single-state dynamics, and results from multi-state fewest switches surface hopping (FSSH) algorithm solvated trajectories. An outline of this paper is as follows.
In Section~\ref{sec:theory}, we discuss the theory of additive QM/MM with electrostatic embedding, nonadiabatic dynamics,  and the implementation choices made in INAQS. 
In Section~\ref{sec:applications}, we demonstrate INAQS' capacities for AIMD and QM/MM umbrella sampling. 
In Section~\ref{sec:FSSH}, we present surface hopping
results.
We conclude in Section~\ref{sec:conclusion}.

\section{Theory}\label{sec:theory}
\subsection{The QM/MM Hamiltonian}\label{sec:theory:qmmm}
In QM/MM schemes, like other hybrid quantum/classical approaches\cite{Tomasi81, Mennucci:2012ct, Curutchet2017}, the total energy of the system, $\mc{S}$, can be written as the sum of the energy of 
the QM inner subsystem $\mc{I}$, $E_{QM}(\mc{I})$, the MM energy of the environment (outer subsystem) $\mc{O}$, $E_{MM}(\mc{O})$, 
and their interaction, $E_{QM-MM}(\mc{I,O})$\cite{Senn:2009gk}
\begin{equation}
    E(\mc{S}) = E_{QM}(\mc{I}) + E_{MM}(\mc{O}) + E_{QM-MM}(\mc{I,O})
    \label{eq:QMMM}
\end{equation}
The MM region is described by a classical force fields generally composed of harmonic bonded interactions (bonds, angles, and dihedrals) and Coulomb and Lennard-Jones interactions for the non-bonded interactions.
The QM region is computed with an appropriately chosen electronic structure method.
In mechanical embedding, all interactions between the QM and the MM region are treated at the MM level of theory and the QM contribution to the total energy is obtained by performing a vacuum calculation on QM region.
Thus, no polarization of the QM system due to the environment takes place.
The electrostatic interactions  between QM and MM region are treated purely classically in terms of the Coulomb interaction of fixed point charges, which can be a problem especially if the system undergoes a chemical reaction that significantly changes its electronic density. 

The missing polarization of the QM region in the mechanical embedding can be addressed by the electrostatic embedding scheme, where the non-bonded electrostatic interactions between QM and MM region are treated at the QM level of theory.
Non-bonded, non-electrostatic interactions (Lennard-Jones) are still computed classically. 

Due to the inclusion of the charges of the MM region into an effective interaction Hamiltonian, the environment is able to polarize the electronic density:
\begin{equation}
    \hat{H}^{eff}\ket{\Psi} = \left(\hat{H}^0 + \hat{H}^{QM-MM}\right) \ket{\Psi} = E\ket{\Psi}
\end{equation}
Here $H^0$ is the Hamiltonian of the isolated QM subsystem, $E_{QM}(\mc{I})$, and $H^{QM-MM}$ is the electrostatic coupling between the inner and outer subsystems.
For the electrostatic embedding this interaction operator is given by an additional nuclear-like 1-electron term in the Hamiltonian:
\begin{equation}
    \hat{H}^{QM-MM} = \sum_k \frac{q_k}{\left| \vec{R}_k-\hat{\vec{r}}_1 \right|}
\end{equation}
where $q_k$ is the charge of the MM atom $k$ at position $\vec{R}_k$. %
When computing the resulting forces on the MM atoms, one simply takes the product of the charge on the MM atom and the electric field arising from the electronic density and QM nuclear charges evaluated at $\vec{R}_k$:
\begin{equation}
    \vec{F}_k{} =  q_k\vec{E}(\vec{R}_k; \rho)
\end{equation}
This approach can be easily generalized to excited states. As one should expect, our interface, which implements an electrostatic embedding, can produce a mechanical one ``for free'' simply by turning off the 1-electron terms and turning the classically computed electrostatics back on.
INAQS does not currently support polarizable embeddings.

\subsection{Excited-State Dynamics}\label{sec:theory:dynamics}
INAQS models the excited state nonadiabatic dynamics of medium to large molecular systems in solvated environment using on-the-fly linear-response Ehrenfest dynamics or fewest switches trajectory surface hopping\cite{tully1990JCP} dynamics. 
Because the physics of Ehrenfest\cite{doltsinis:2002:review} and surface hopping\cite{truhlar:review:surfacehop} dynamics are well described in the literature, here we will describe the theory behind these methods only briefly, before describing in detail the practical numerical issues that go along with the INAQS implementations.

For both linear reponse Ehrenfest and surface hopping dynamics, 
the nuclear and electronic degrees of freedom are propagated separately.
Propagation of the electronic degrees of freedom is straightforward.
The electronic degrees of freedom are represented by a time-dependent electronic wavefunction $\Psi^{\textrm{elec}}$, expanded in a known set of (time-dependent) basis functions $\Phi_i$,
\begin{equation}
    \Psi^{\textrm{elec}}(t) = \sum_i c_i(t) \Phi_i \qty(\vec{R}(t)).
\end{equation}
Here, $\vec{R}$ is the vector of nuclear positions.
The set $\left\{\Phi_i(\vec{R}(t))\right\}$ are typically selected as the wavefunctions of the adiabatic states of the system and the $c_i(t)$ denote the corresponding time-dependent weights of each state $i$.
The electronic wavefunction $\Psi^{\textrm{elec}}$ is propagated using the time-dependent Schr\"{o}dinger equation,
\begin{equation}
    i\hbar \dfrac{\partial}{\partial t} \Psi^{\textrm{elec}}(t) = \hat{H}^{\textrm{elec}} \Psi^{\textrm{elec}}(t)
\end{equation}
with $\hat{H}^{\textrm{elec}}$ being the electronic Hamiltonian of the system and $\hbar$ begin the reduced Planck constant.
In a moving basis (e.g. the adiabatic basis), this equation reduces to a standard set of coupled equations for the expansion coefficients $c_i(t)$,
\begin{equation}\label{eq:cit}
  i\hbar \dot{c}_i = \sum_j c_j \qty[V_{ij} - i\hbar\dot{\vec{R}}\cdot \vec{d}_{ij}]
\end{equation}
where the $V_{ij} = \mel{\Phi_i}{\hat{H}^{\textrm{elec}}}{\Phi_j}$ are the matrix elements of the electronic Hamiltonian, and $\vec{d}_{ij}=\mel{\Phi_i}{\frac{\partial}{\partial \vec{R}}}{\Phi_j}$ are the derivative couplings between adiabatic states ${\Phi_i}$ and ${\Phi_j}$.
Within INAQS, \refeq{eq:cit} is integrated via matrix exponentiation in order to propagate the expansion coefficients from $t$ to $t+dt$.

This completes the straightforward electronic propagation. The propagation of the nuclei is more demanding and requires a strong semiclassical approximation, one based either on mean-field Ehrenfest dynamics or on  state-specific surface hopping dynamics. In general, 
the nuclei follow Newton's equation of motion:
\begin{equation}\label{eq:newton}
    \mat{M} \ddot{\vec{R}} = - \vec{\nabla} E_{\lambda}
\end{equation}
where $M$ is the diagonal matrix of masses of all nuclei, $E_{\lambda}$ is the energy of an electronic state $\lambda$. The key point is the definition of $\lambda$?

\subsubsection{Ehrenfest}
Ehrenfest dynamics is  a mean-field theory that is most accurate when $(i)$ the adiabatic surfaces of interest are nearly parallel (so that moving along an average surfaces makes a lot of sense)\cite{bellonzi:2016:sqc}; or $(ii)$ when the nuclear motion is fast and there is less separation between nuclear and electronic time scales (an extreme case being classical photon - quantum electron interactions\cite{theta:2019:ehr1}).
Over the years, a host of nonadiabatic dynamics based on the Ehrenfest equations of motion (but improved upon by using different quasi-classical initialization schemes) have been developed.\cite{meyer_miller:1979,kapral:2008:jcp_pbme,miller:2014:sqc} For now, INAQS has implemented only standard Ehrenfest dynamics based on simple classical sampling of the initial conditions.

Mathematically, for Ehrenfest dynamics, the effective energy is the average energy:
\begin{eqnarray}
E_{\lambda} = \sum_{ij} {c_i^*  V_{ij} c_j}
\end{eqnarray}
One must be careful when differentiating this expression. For linear response Ehrenfest dynamics expressed in an adiabatic basis of electronic states, the correct force is:
\begin{equation}\label{eq:lrehrenfest}
    \ddot{\vec{R}} = - \mat{M}^{-1} \left( \sum_i {\norm{c_i}}^2 \vec{\nabla} V_{ii} - \sum_{ij} c_i^*  c_j \vec{d}_{ij} (V_{jj} - V_{ii})\right)
\end{equation}
where we note that the last term in parentheses is real-valued and equivalent to $\sum_{ij} \Re\left[c_i^*c_j\right]  \vec{d}_{ij} (V_{jj} - V_{ii})$.

\subsubsection{Surface Hopping}
Surface hopping takes a different approach from Ehrenfest dynamics and propagates nuclei on one ``active'' adiabatic surface (with the possibility of hops to another adiabatic surface).
Because nuclei usually move far slower than electrons, for molecular systems, trajectory surface hopping usually has a greater regime of applicability than Ehrenfest dynamics. The former method also recovers detailed balance\cite{tully:2005:detailedbalance,tully:2008:detailedbalance} unlike the latter (though some advances have been made recently \cite{miller:2015:jcp_detailedbalance}).

There are many subtleties associated with the implementation of surface hopping in practice.
The implementation in INAQS largely follows Jain and coworkers\cite{jain2016:afssh} with some modifications.
The salient features are as follows:
\begin{enumerate}
  \item The wavefunction is propagated following \refeq{eq:cit} via an extended  Meek and Levine overlap scheme\cite{meek2014:dcevaluation,jain2016:afssh} that significantly extends the time step from the original finite difference approach suggested by Hammes-Schiffer and Tully\cite{hammesschiffer1994:protontransfer,barbatti2009:overlaps}.
  In brief, we take 
  $$\dot{\vec{R}}\cdot \vec{d}_{ij} = \frac{1}{dt}{\left(\log \mat{U}\right)}_{ij} $$ where 
  $(\mat{U})_{ij} = \braket*{\Phi_i(t)}{\Phi_j(t+dt)}$.
  According to this expression, we effectively average the derivative coupling term over the duration of the entire classical time step.
  Maintaining the orthogonality while computing the matrix logarithm of a unitary matrix is nontrivial and thus we employ a technique based on the Schur decomposition.\cite{loring2014:matlog}
  With this simplification, derivative couplings need \emph{not} be computed at every time step, but rather only when there is a transition.
  Note that in Fig~\ref{fig:inaqsflow} the computationally less demanding right ``no'' branch is most often followed afterwards by  `Need to hop?'.
  Highly efficient, exact matrix overlaps for spin-flip CIS and TD-DFT states have recently been implemented in Q-Chem.\cite{chen2022:cisoverlaps}
  
  \item Since the phases of the wavefunctions are undefined, the signs of the columns of $\mat{U}$ are also formally undefined.  For smooth dynamics, one can always choose the diagonal elements to be positive (so-called ``parallel transport''); however, in the presence of trivial crossings\cite{zhou2020:overlaps,lee2019:trivialcrossing} (particularly sharp crossings where the diabatic coupling is nearly zero and a hop guaranteed) such a scheme is ill-behaved. To that end, INAQS implements the protocol outlined by Zhou and coworkers\cite{zhou2020:overlaps} that aims to pick adiabatic signs by minimizing a surrogate for the function $\Tr[\abs{\log \mat{U}}^2] \approx \Tr[3\mat{U}^2-16\mat{U}]$.

  \item For velocity reversal, we follow Jasper and Truhlar\cite{truhlar2003:reversal}, reversing the nuclear velocity along the direction of the derivative coupling whenever a hop fails and $-\vec{\nabla} V_{j} \cdot \vec{d}_{\lambda j} \vec{d}_{\lambda j}^{\;T} \mat{M}^{-1} \dot{\vec{R}}  < 0$; \emph{i.e.} when the momentum projected along the derivative coupling opposes the force from the surface the system failed to reach. This protocol was found to be superior \cite{jain2020:privcomm_reversal} to that suggested in Ref.\citenum{jain2016:afssh}.
  
  \item Decoherence certainly can play an essential role in surface hopping.  For instance, it is known that, in some regions, surface hopping recovers incorrect scaling laws for the Marcus problem\cite{landry:2011:marcus_fssh,reichman:2016:fssh}. Nonadiabatic transition state theory can also suffer without decoherence\cite{jain:2015:jpcl}.  INAQS has implemented a module for decoherence, and we will report the effects and necessity of decoherence effects in the condensed phase in a later publication.
\end{enumerate}

\subsection{Code Infrastructure}

Born-Oppenheimer, Ehrenfest and surface hopping dynamics all  propagate Newton's equations and conserve energy.
In developing INAQS, we sought to exploit this conservation law to develop an interface with minimal intrusion on the MD driver.
In practice, trajectory-based MD simulations can be crudely divided into two parts:
$(i)$ A nuclear propagator that determines the next position of the nuclei in time by integrating a given force and $(ii)$ a force provider that computes molecular gradients for the given nuclear (and possibly electronic) configuration.

Let us first consider the role of the force provider for each specific dynamics scheme considered above.
In the case of purely classical MD, the gradient is obtained \emph{via} (pre-determined) classical force fields, typically representing the ground electronic state.
For single-state Born-Oppenheimer dynamics, the force provider computes a gradient for a fixed electronic state.
For Ehrenfest dynamics or surface hopping, however,  the forces depend on the excited electronic states and for each state there is a different gradient. For Ehrenfest dynamics, the force provider must pass a gradient that incorporates knowledge of the corresponding electronic wavefunction and (for linear-response as opposed to real-time Ehrenfest) also the nonadiabatic couplings between the electronic states (see \refeq{eq:lrehrenfest}).  For surface hopping dynamics, the provider must pass an electronic state gradient that incorporates all hopping information.

Second, let us address the nuclear propagator within the MD code.
The nuclear propagator receives the forces and  integrate Newton's equations (\refeq{eq:newton}) to determine the new positions and momenta.
In general,  the nuclear propagator does not care how the forces are obtained---whether from force fields (molecular mechanics), an electronic structure calculation, or a hybrid QM/MM setup.
One may even apply \emph{ab initio} exciton or fragment-based Hamiltonians \cite{Sisto2014, Sisto2017, Menger2018, persico2021}.
Thus, in general, a call to update the nuclear position
should be valid across a variety of different algorithms from classical MD to \emph{ab initio} MD to Ehrenfest dynamics and surface hopping.
There is  one difference between surface hopping and the other algorithms presented above; following Pechukas\cite{pechukas:1969:pr} and Herman\cite{herman:1984:jcp:rescaling_direction}, in the case of a hop, the velocities of the system need to be rescaled to conserve the total energy. Even so, a carefully designed interface can be quite simple and very general.
The decoupled nature of the  nuclear and electronic propagation schemes allows a variety of trajectory-based MD approaches to be implemented by way of the same code paths.

\begin{figure}
    \centering
    \includegraphics[width=0.87\textwidth]{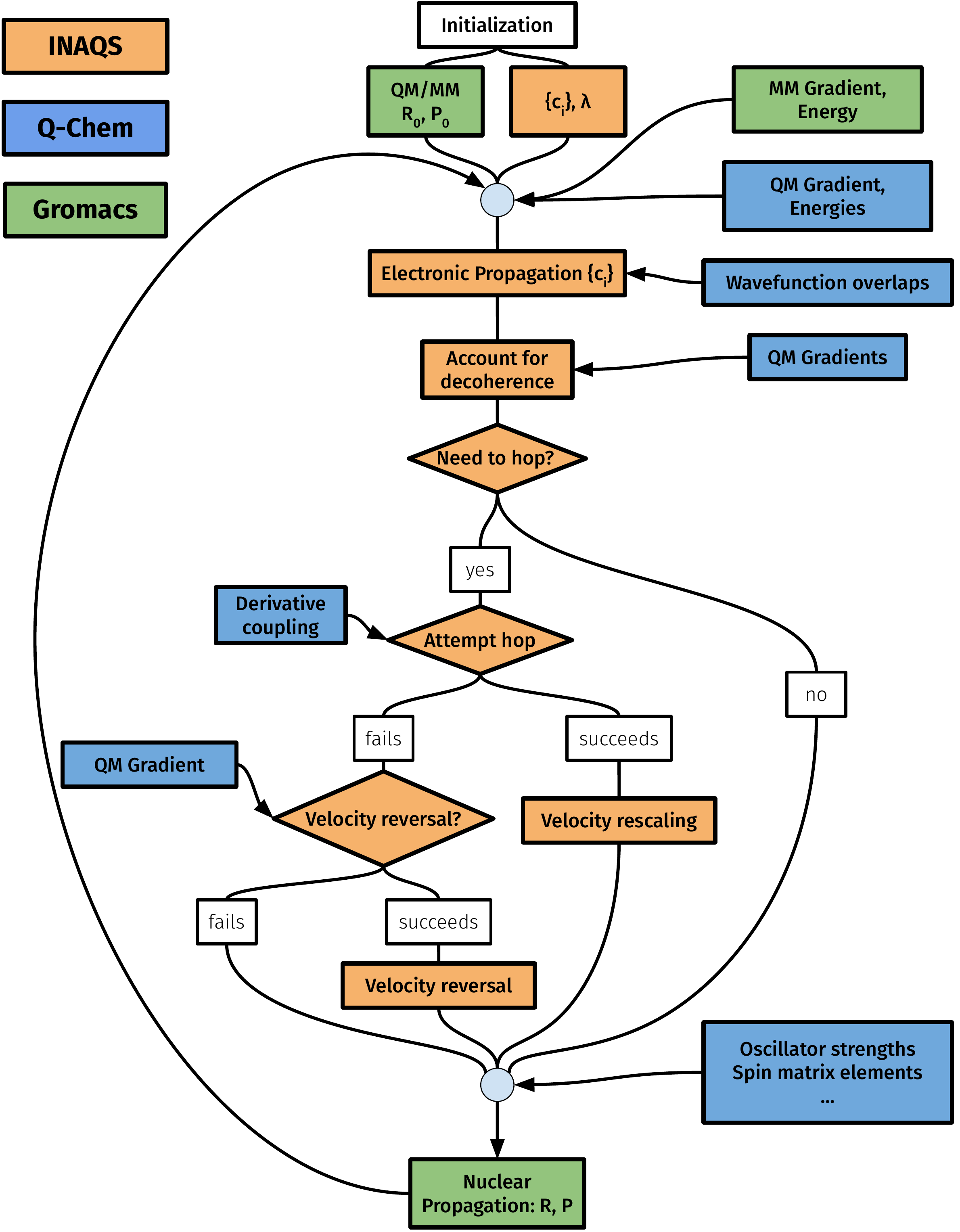}
    \caption{Fundamental steps of a surface hopping molecular dynamics simulation. INAQS (in tan/orange) acts as an interface between the electronic structure software (Q-Chem, in blue) and the MD driver (Gromacs, in green) to track and integrate the electronic wavefunction, provide relevant electronic properties of the system, and effect velocity rescaling or reversal during a hopping event. INAQS hooks into Gromacs at exactly 2 points: once to supply QM energies and gradients (as is typical in an MD code) and again to optionally update velocities during a hopping event. The code is quite general and we support single-state calculations or Ehrenfest dynamics via the same codepaths.
    }
    \label{fig:inaqsflow}
\end{figure}

In practice, implementing surface hopping is more difficult than Ehrenfest dynamics, and so we will focus on the former here.  The basic scheme for surface hopping is presented in \fig{fig:inaqsflow}.
The INAQS software implements function calls in two places within the GROMACS package.

\subsubsection{GROMACS interface}
The current implementation of INAQS is tied to a modified version of the Gromacs, which implies that one requires all of the usual inputs as appropriate for a classical MD simulation.
The INAQS and GROMACS codes are linked together at the source level to allow direct memory access for communication.
Like many MD codes\cite{allentildesley1989}, Gromacs preferentially uses a leapfrog integrator.
Such an integrator is not usable for surface hopping because velocities and positions are not obtained for the same time, but rather interleaved.
velocity-Verlet is available within Gromacs, but is implemented using the same code-paths as leapfrog.

INAQS requires only 2 function calls be inserted into Gromacs (or any other MD program):
\begin{enumerate}
    \item The first call computes QM energies and forces.
    \item The second call updates velocities and the gradient in the event of a hop.
\end{enumerate}

To understand where these two calls are placed exactly, consider the structure of the existing GROMACS integrator.
At the start, a force is required to propagate the current nuclear coordinates
\begin{equation}
    \vec{F}(t) = {\left.-\vec{\nabla} V\right|}_{\vec{R}(t)}.
\end{equation}
A standard Gromacs code computes the MM forces and Q-Chem the QM energies, forces, and any other requested properties. {\bf First call:} INAQS constructs the necessary input for Q-Chem and returns the results via Gromacs' additive QM/MM routines to update the force. During this time, INAQS also integrates \refeq{eq:cit} for the electronic coefficients and determines if a hop is necessary. 

The next step in the Gromacs' MD routine is the first velocity half-step:
\begin{equation}\label{eq:vvhalfstep1}
    \vec{v}(t) = \vec{v}(t-\frac{\Delta t}{2}) + \mat{M}^{-1}\vec{F}(t)\frac{\Delta t}{2}.
\end{equation}
{\bf Second call:} After \refeq{eq:vvhalfstep1}, INAQS determines whether or not a hop should occur and implements velocity rescaling and/or reversal.  If the code determines that a hop succeeds, we modify the velocities using the usual energy conservation equations, and then update Gromacs' QM potential energy and gradient.

At this point, Gromacs enforces any applicable constraints via SHAKE, RATTLE, and/or LINCS; note that, having adjusted the potentials and velocities {\em beforehand}, INAQS  is compatible with such constraints. 

Finally, Gromacs takes its second velocity half-step:
\begin{equation}\label{eq:vvhalfstep2}
    \vec{v}(t+\frac{\Delta t}{2}) = \vec{v}(t) + \mat{M}^{-1}\vec{F}(t)\frac{\Delta t}{2}
\end{equation}
and updates the nuclear positions:
\begin{equation}
    \vec{x}(t+ \Delta t) = \vec{x}(t) + \vec{v}(t+\frac{\Delta t}{2}) \Delta t
\end{equation}

The INAQS repository contains a locally modified version of Gromacs 4.6.5 implementing these changes.
INAQS is compatible with
enhanced sampling protocols like umbrella sampling via Gromacs's native algorithms or PLUMED2\cite{plumed2} (which also connects to Gromacs).
INAQS uses atomic units exclusively internally; conversion to and from the MD code's unit system is performed automatically.

\subsubsection{Q-CHEM implementation}
While INAQS is relatively decoupled from Gromacs on the MD side, the interface to Q-Chem is much more extensive.
This extra coupling is a result of the fact that many more choices must be communicated to the electronic structure  software (basis sets, functionals, convergence algorithms) as opposed to the MD software (which, by design, mostly handles itself).
Communication with Q-Chem is handled via a system call rather than direct linking, but INAQS directly reads the binary intermediaries (rather than parse any ASCII output).
The Q-Chem execution environment is established in the usual way with Q-Chem's standard environmental variables, indicating the location of the executable and scratch directories.
INAQS allows the user to control the number of threads used for parallel execution; see Appendix~\ref{sec:usage} for an example of an input file.
Calls to Q-Chem are heavily optimized to avoid recalculation of the SCF or excitation amplitudes while maintaining flexibility in the kinds of properties that we compute.
In practice, we find that per geometry, INAQS implementation is 2-3 times faster than naive job submission would be; see Appendix~\ref{sec:computation} for explicit performance data.
INAQS is compatible with Q-Chem versions 5.4 and higher in their unmodified state.
Spin-flip overlaps for spin-flip nonadiabatic dynamics 
are implemented starting in Q-Chem version 6.0.

\subsubsection{Availability}
INAQS is written in the C++11 programming language and provides an easy to use C  (specifically C89) interface, allowing compatibility with almost all programming languages and existing software packages.
The source code is available on GitHub at \url{github.com/INAQS/inaqs}.
INAQS is written in a modular way such that the code can be easily adapted to another MD package that uses the velocity-Verlet algorithm.
In principle, one can also change electronic structure software so long as the new package can evaluate the overlap of electronic wavefunctions.

\section{Applications}\label{sec:applications}

\begin{figure}
    \centering
    \raisebox{-0.5\height}{\includegraphics[width=0.5\textwidth]{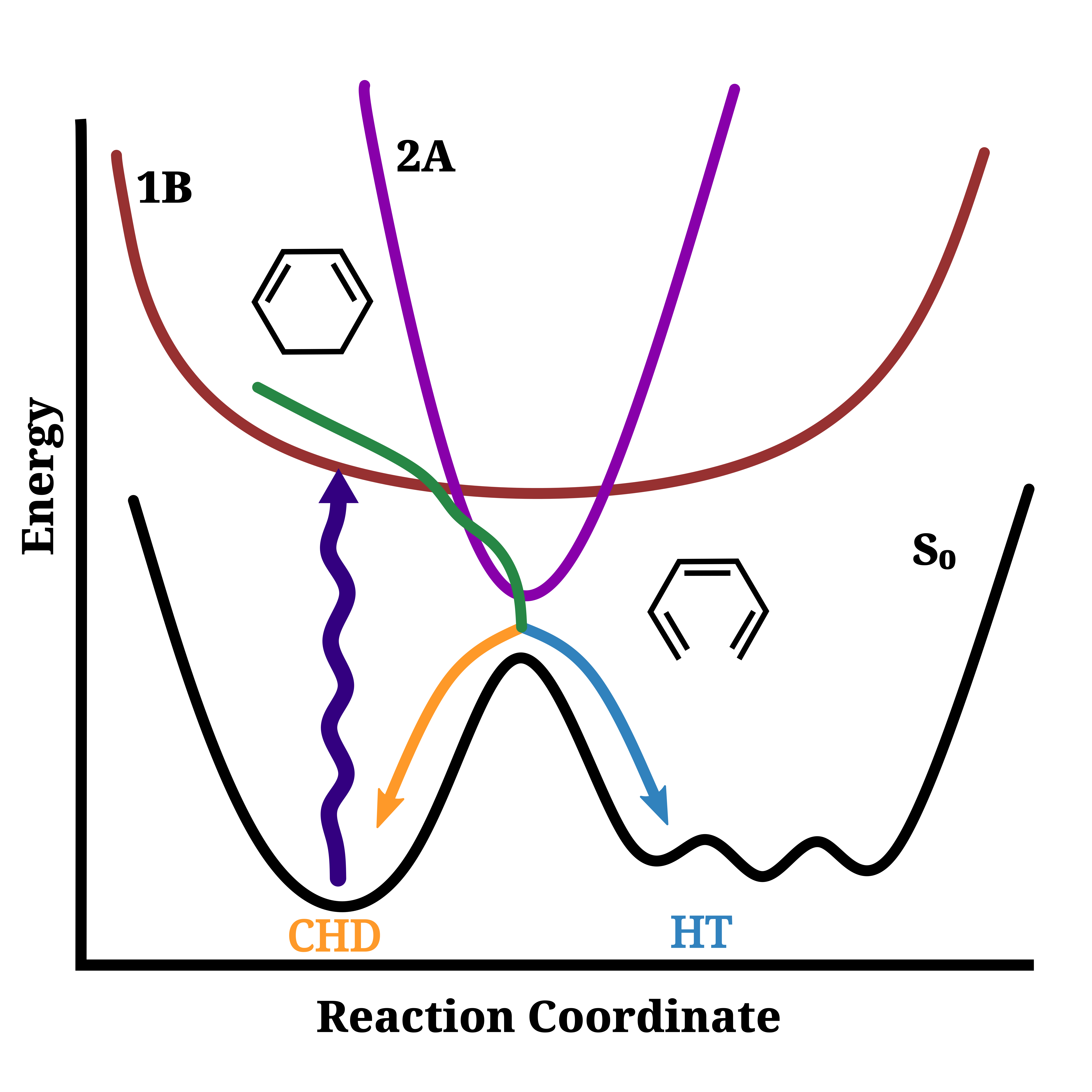}}
    \raisebox{-0.5\height}{\includegraphics[width=0.45\textwidth]{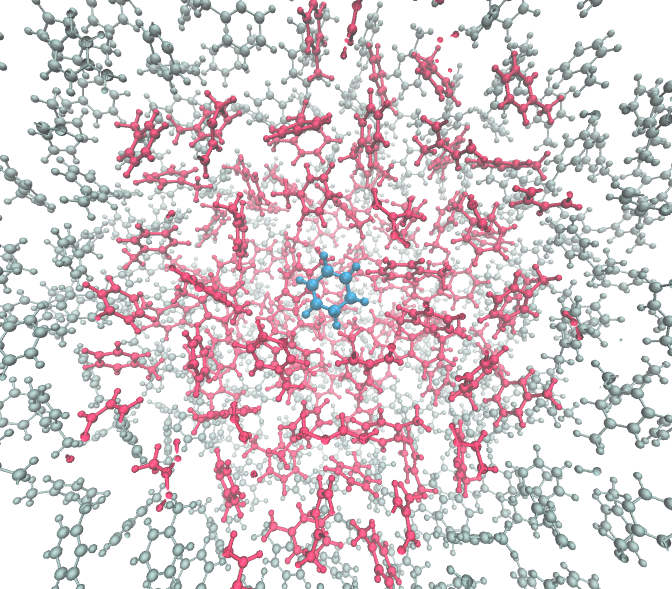}}
     \caption{\label{fig:CHDscheme}Schematic of the photo excitation and relaxation process for CHD.}
\end{figure}

The photochemical interconversion between 1,3-cyclohexadiene (CHD) and hexatriene (HT) is a common example for a (4n+2) photo electrocyclic reaction, following the Woodward-Hoffmann rules
and has been intensively studied\cite{deb2011:CHDReview} both experimentally\cite{trulson1987femtosecond,rudakov2009ground,weber2006:CHD} and theoretically\cite{garavelli2001:chd,xing1999:chd,tamura2006:CHDdynamics}. 
In general, there is widespread interest in the structure and dynamics of CHD and substituted
CHD, as this class of molecules plays a crucial role in many biological processes---e.g. the photobiological synthesis of vitamin D3---and also is a target for platform for the design of molecular photoswitches.
That being said, photochemistry and photoswitches do not operate in vaccuum, but rather in complex environments (e.g., solvent) and it is essential to account for solvent if one seeks an accurate description of such processes.
In the following, we will present: $(i)$  the absorption spectrum of CHD in different solvents (toluene and ethanol) as computed by Born-Oppenheimer AIMD trajectories; $(ii)$ 
the free energy profile of the ring opening reaction, i.e. CHD to HT conversion, as computed by  umbrella sampling; $(iii)$ 
photochemical dynamics of the excited state ring opening as computed by surface hopping calculations.  Most of these applications are not new \cite{NewtonxQMMM,groenhof2015:qmmmPMF,turbomole} \emph{per se}, but are offered as a means of highlighting the capabilities of INAQS.

\subsection{Absorption spectra}\label{sec:spectrum}
The photo-absorption spectrum of the CHD is simulated using vertical excitation energy calculations obtained from three different approaches; $(i)$ single point calculation of the ground state equilibrium with implicit solvent model (CPCM), $(ii)$ 2000 snapshots taken from a 2~ns classical NVT MD simulations with explicit inclusion of the solvents (ethanol or toluene molecules), $(iii)$ $2\cdot10^4$ structures obtained from a 20~ps ground state AIMD simulation via electrostatic embedding. To investigate the extent to which the absorption spectra are influenced by the presence of the implicit and/or explicit solvent models, we also simulated the absorption spectrum for all three approaches in vacuum. Absorption spectra were obtained as a normalized superposition of Gaussians localized at computed excitation energies ($\varepsilon_i$),
\begin{equation}
  \label{eqn:spectrum}
  I(\varepsilon) \propto \sum_{i} f_i / \sigma \cdot \exp \left[-\dfrac{(\varepsilon - {\varepsilon_i)^2}}{\sigma^2} \right]
\end{equation}
with $f_i$ being the oscillator strength, $\sigma$ the spectral broadening. %
Spectra are normalized to have a constant absorption cross section, \emph{i.e.} $\int_{-\infty}^{\infty}  I(\varepsilon)~d\varepsilon = 1$. The broadening, $\sigma$, is chosen to be  0.1~$eV$ for $(ii-iii)$ and 0.3~$eV$ for approach $(i)$; the latter broadening parameter must be larger than the former in order for the different methods to match; after all, only the former includes inhomogeneous broadening.  

\begin{figure}
    \centering
     \includegraphics[width=\textwidth]{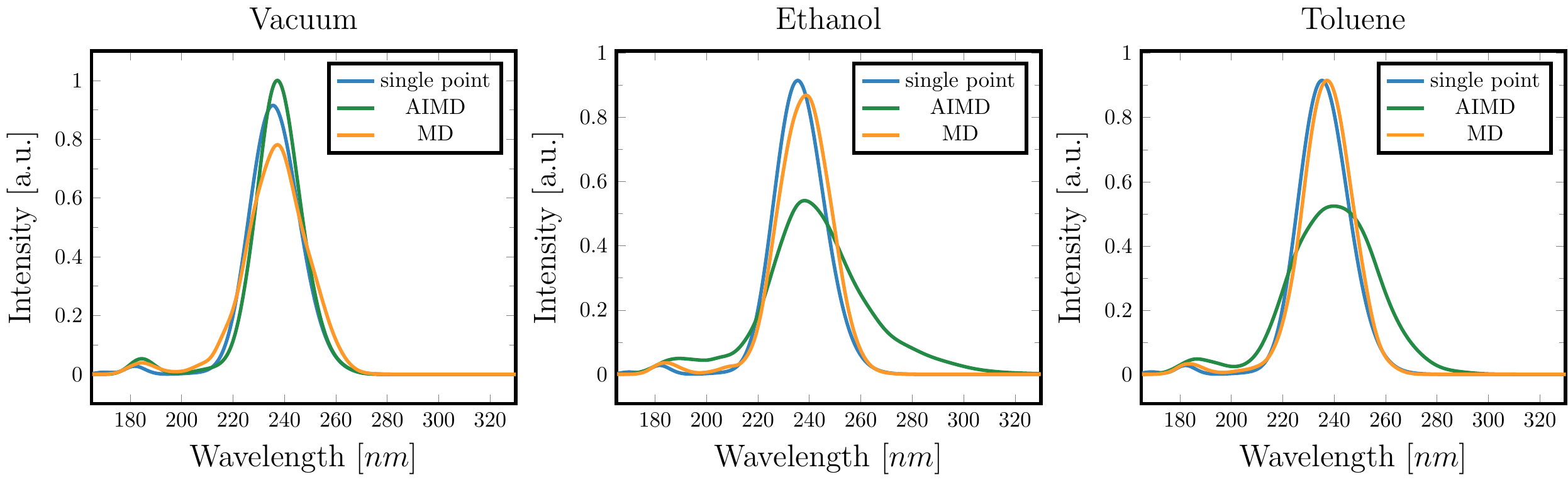}
    \caption{Absorption spectrum of the CHD simulated from $(i)$ a single point calculation of the ground state equilibrium with a CPCM implicit solvent model (single point, blue), $(ii)$ snapshots taken from classical MD simulations with explicit inclusion of the solvents (MD, orange), $(iii)$ snapshots from structures obtained from ground state AIMD (AIMD, green) via electrostatic embedding.
    }
    \label{fig:spectrum}
\end{figure}

In all the three approaches, the spectrum is dominated by the first bright excited state with an excitation energy of around $5.3~eV$. As it is evident from \fig{fig:spectrum}, the inclusion of the solvent has no significant effect on the position of the peak maximum ($\approx~0.05~eV$). This lack of change should not be surprising, as CHD is a small rigid molecule that oscillates mainly around its ground state equilibrium structure. Note that the classical, QM-derived force field does capture the correct ground state potential energy landscape for the MD simulation of the isolated CHD. However, the method fails to capture the polarization of the QM region in the presence of the explicit solvents. This failure can be seen by comparing the spectra obtained using ground state AIMD and MD for isolated CHD and solvated CHD in \fig{fig:spectrum}; the spectrum obtained via ground state AIMD results in a significantly broader spectrum when the explicit solvents are present (with toluene showing a slightly larger broadening), while both methods result in approximately the same spectrum for the isolated CHD. 

The observations above lead us to hypothesize that the presence of solvent mainly modifies the \emph{ground state} potential energy landscape, while the vertical excitation energies are only marginally effected within electrostatic embedding. To further confirm this hypothesis, we calculated the RMSD for the excitation energies computed with and without the explicit solvent for 2000 structures, which is found to be  $\approx0.01~eV$ for the first 10 excited states. However, the energetic fluctuations due to structural changes obtained from calculations with explicit solvent are an order of magnitude larger ($\approx0.2~eV$). Thus, we conclude that the spectral broadening in the AIMD of solvated CHD is mainly attributed to structural fluctuations in the ground state.

\subsection{Umbrella sampling}
Using INAQS, we have modeled the thermally-induced ring-opening reaction (CHD to HT) in the ground state (see \fig{fig:CHDscheme}) using umbrella sampling\cite{chandler:statmech, gwham}. The free energy landscape is computed along the ring-opening reaction coordinate within an electrostatic QM/MM embedding (see Appendix~\ref{sec:computation} for computational details). 

\begin{figure}
    \centering
        \includegraphics[width=0.9\textwidth]{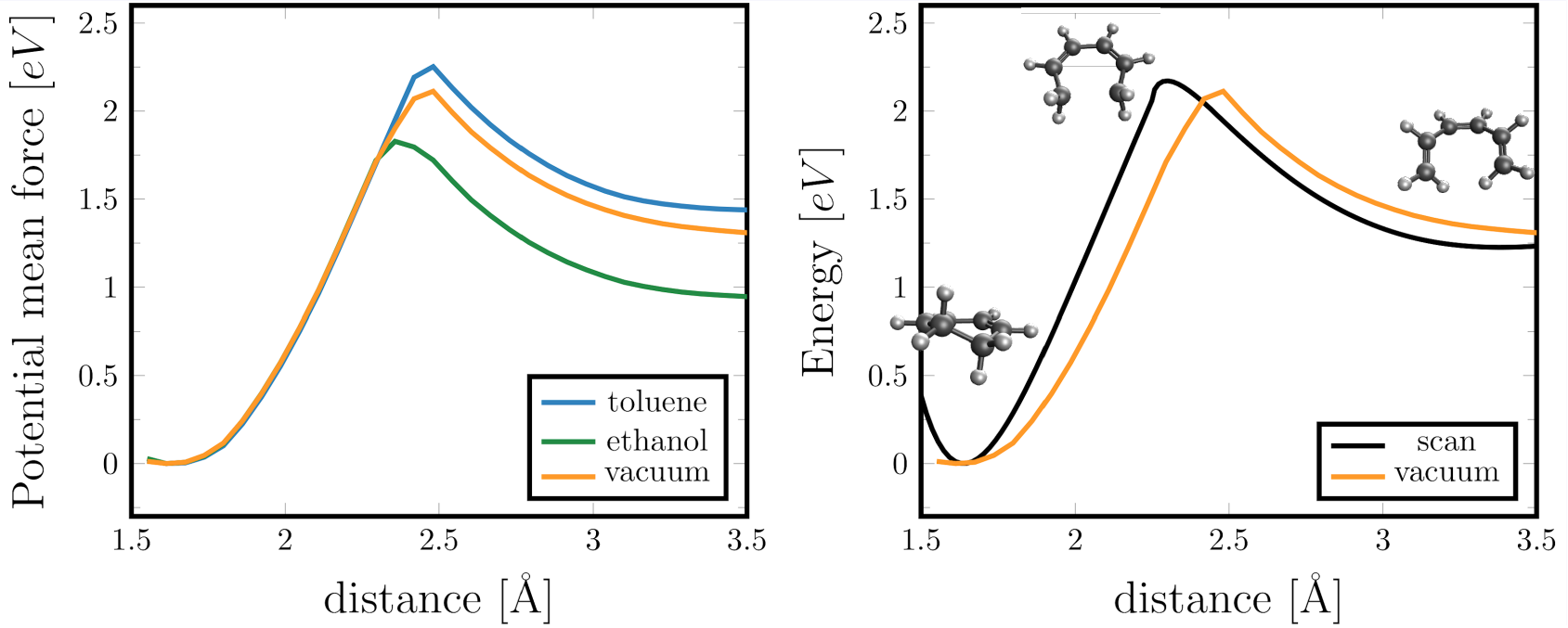}
    \caption{(left) Potential of mean force along the reaction coordinate of the ring opening of CHD in different environments toluene (blue), ethanol (green) and the isolated molecule (orange). (right) Vacuum comparison between the potential of mean force (orange) and the relaxed surface scan using Q-Chem (black) for CHD. %
    }
    \label{fig:umbrella}
\end{figure}

The distance between the center of mass (COM) of the two \ce{CH_2} groups of the CHD is chosen as the pulling coordinate. First a single  pull is performed to sample the initial structures for the windows of the umbrella sampling; a 10~ps MD run is performed with a pulling rate of 0.5~\AA/ps along the ring-opening coordinate ranging from 1.5~\AA\ (closed) to 3.5~\AA\ (open).   For this trajectory, the molecule is driven quickly enough such that the environment may not have time to fully relax. Second and subsequently, a 2~ps MD simulation with fixed harmonic constraint for the reaction coordinate was performed in 0.1~\AA-wide windows starting from three selected structures from the initial pulling simulation. The free energy profile for different solvents are shown in \fig{fig:umbrella} and compared to the potential energy curve obtained by a relaxed surface scan of the isolated CHD using Q-Chem along the C-C bond distance. In all cases, the maximum of the potential is found to be at around $2.4\pm0.1$~\AA\ of the \ce{CH_2}-\ce{CH_2} COM distance, with toluene showing the smallest distance, ethanol the largest, and isolated CHD being in the middle. It is worth mentioning that the barrier in ethanol is lowered by 0.5~eV and the open ring form (HT) is stabilized by 0.5~eV compared to the toluene case. The energy profile obtained using umbrella sampling of the isolated CHD agrees well with the relaxed surface scan (subfigure on the right). The high energy barrier observed in the ground state (more than $\approx42$ Kcal/mol $=71 k_{\textrm{B}} T$) prevents thermal interconversion between the closed-ring and open-ring forms.

\subsection{Surface Hopping}\label{sec:FSSH}
There is an extensive experimental\cite{weber2006:CHD} and theoretical\cite{tamura2006:CHDdynamics} literature\cite{deb2011:CHDReview} exploring the  photo-induced ring opening dynamics of CHD. The basic physics is that the ground state functions as a double well with CHD and HT as two stable isomers. Upon photoexcitation, there are two excited states of interest: a singly excited state (often referred to as 1B) and doubly excited state (often referred to as 2A). Conical intersections can be identified between the 1B and 2A excited states, as well as between the 2A and $S_0$ ground state. For a schematic figure, see \fig{fig:CHDscheme}.

As described in Appendix~\ref{sec:computation}, we have now run surface hopping calculations at the level of spin-flip TD-DFT for CHD.  While our calculations do not resolve a 1B-2A conical intersection (and the calculations do suffer from spin-contamination\cite{herbert2015:sasf}), we do resolve a strong transition (likely a conical intersection) %
when we monitor the transition from the first excited putative singlet state, $S_1$, to the ground state. In principle, one goal of QM/MM dynamics is to identify the molecular characteristics that guide a reaction to form either  HT ($S_0$-HT) or relax back to CHD ($S_0$-CHD) in solution.

\begin{figure}
    \centering
    \includegraphics[width=0.32\textwidth]{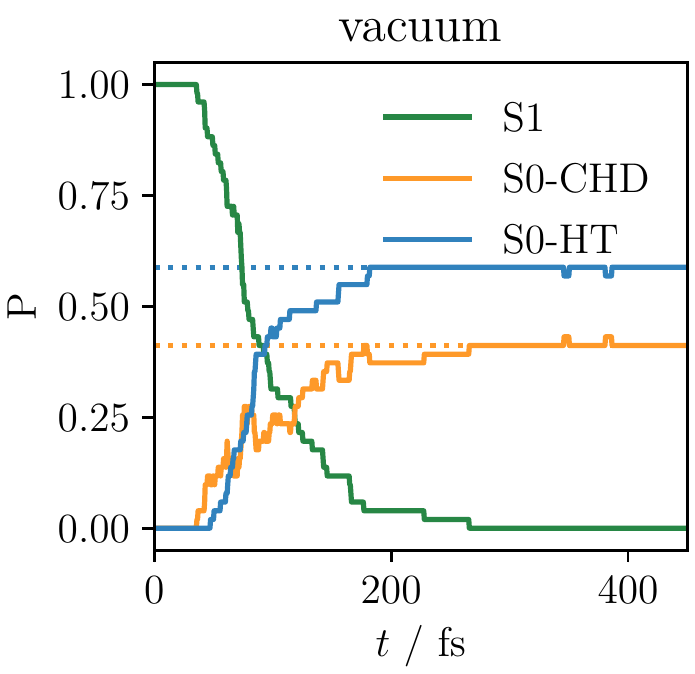}
    \includegraphics[width=0.32\textwidth]{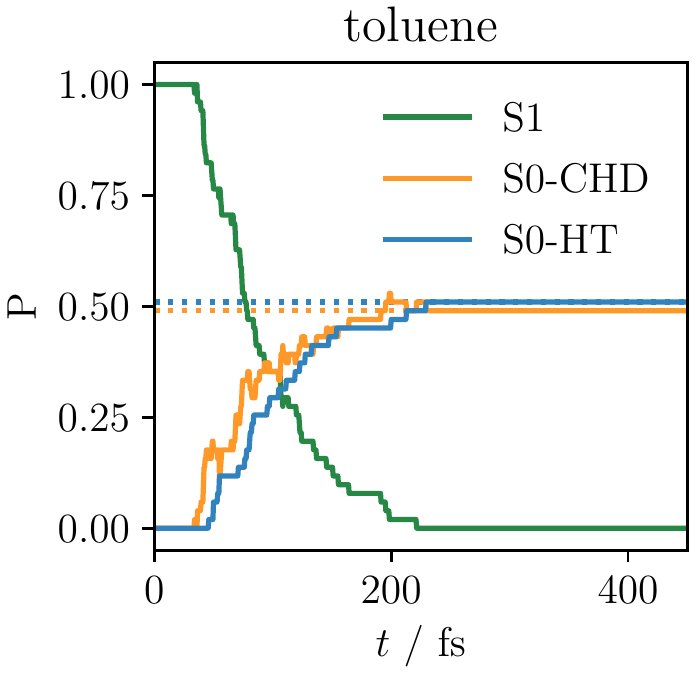}
    \includegraphics[width=0.32\textwidth]{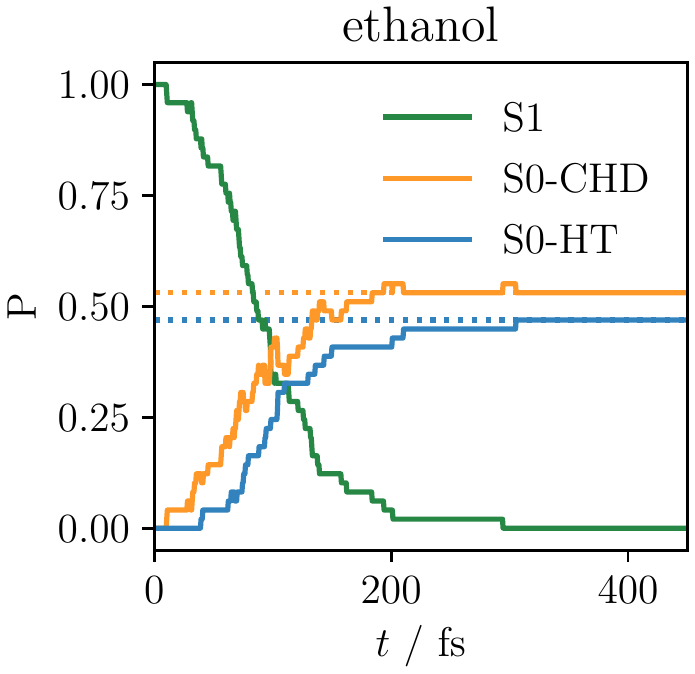}
    \caption{
      The evolution of the populations of the electronic states during surface hopping dynamics for the three simulations in vacuum, in toluene and in ethanol.
      The $S_1$ state (green) rapidly depopulates transfers to the ground state $S_0$, where two different products will be formed, namely the closed-ringed cyclohexadiene (CHD) (orange) and open hexatriene (HT) (blue). Initial excited state-structures were drawn from ground state AIMD simulations as described in Appendix~\ref{sec:computation}.
      \label{fig:populations}
    }
\end{figure}

In \fig{fig:populations}, we plot the population transfer for CHD simulations in three different environments: the isolated molecule (vacuum), in toluene, and in  ethanol. The overall population transfer is quite similar in all cases, where an ultrafast (within the first 200~fs) population transfer from the $S_1$ state to the electronic ground state $S_0$ can be observed. For the first $\approx30$~fs no transition occurs; apparently, this is the length of time needed to reach the coupling region.  
For the two solvent modules (toluene, ethanol) the populations of $S_0$-HT and $S_0$-CHD are nearly the same, with a slightly higher population of the $S_0$-HT state after the population transfer is finished. In both cases a smooth transfer from $S_1$ to $S_0$ can be observed.
In a solution of pentane, it is generally thought\cite{deb2011:CHDReview,havinga1973:CHDyield} that
there is  41\% conversion of CHD to HT; in vacuum, 
experimental indicate are that the yield of hexatriene is nearly unity\cite{deb2011:CHDReview}.
The calculations in \fig{fig:populations} cannot recover these observations quantitatively---the yield of HT in vacuum being 59\% and the solvated yields being less: 51\% in toluene and 47\% in ethanol. In other words, we do recover the correct trends, but we are off quantitatively. The fact that our vacuum calculations do not match experiment indicates %
that the problem must involve more than the QM/MM solvent environment; for example, spin-flip TDDFT is known to suffer several problems as far as reproducing excited-state crossings and barriers quantitatively\cite{casanova2008:sfxcis}. Alternatively, there is always the question of whether or not we should be sampling a Wigner (rather than Boltzmann) distribution with our initial conditions. In any event, qualitatively (though clearly not quantitatively), we do see the expected trend with solvent: the presence of toluene and/or ethanol reduces the yield of HT.

\begin{figure}
    \centering
    \begin{overpic} %
        [width=0.45\textwidth]{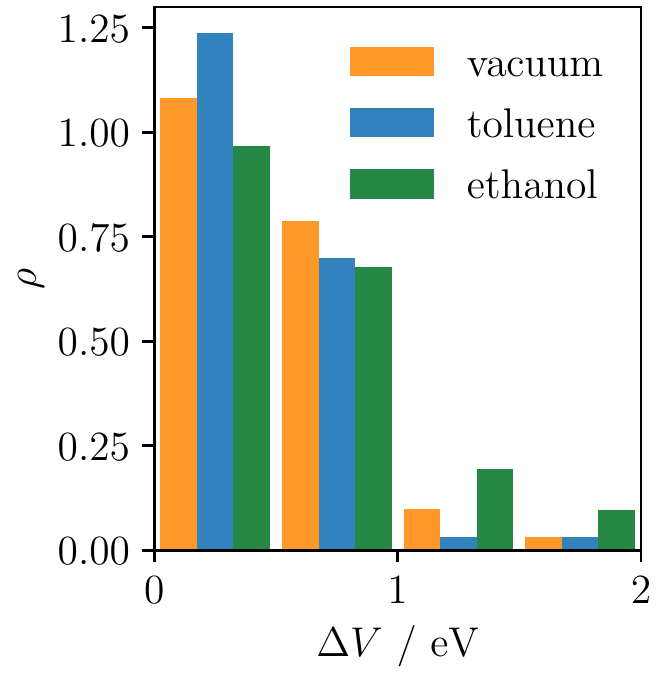}
        \put (-5,90) {\large (a)}
    \end{overpic}
    \hspace{2ex}
    \begin{overpic}
        [width=0.45\textwidth]{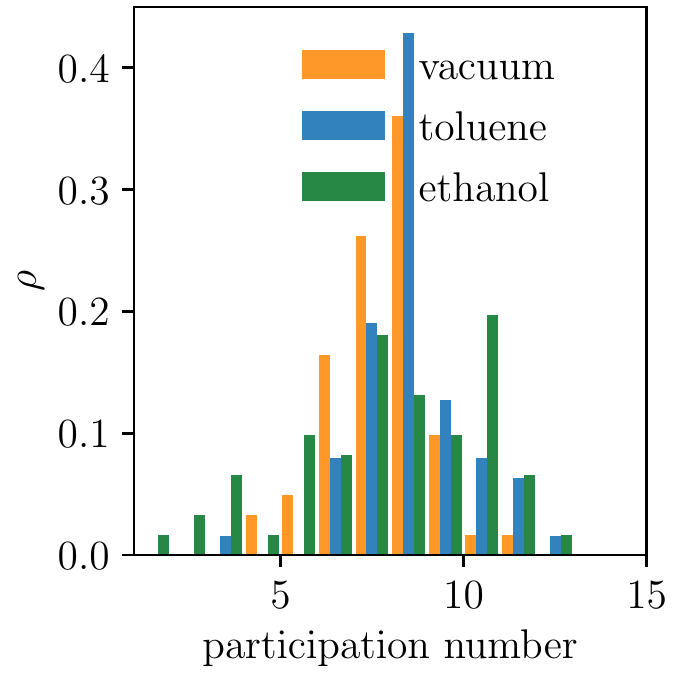}
        \put (-2,94) {\large (b)}
    \end{overpic}

    \caption{a) Distribution of successful (downward) hops by the energy gap and b) participation number for derivative couplings between the $S_1$ and $S_0$ state for the three simulations in vacuum, in toluene and in ethanol. As expected a hop is more likely to succeed if the energy gap is relatively small. For successful hops, the distribution of participation number is modestly broader in solvent than in vacuum.}
    \label{fig:gappartsuccess}
\end{figure}

At this point, one would like to understand how the presence of solvent affects the dynamics.  As discussed above, the solvent alters the potential energy landscape for the ground state but does not greatly affect the relative excitation energies.  Beyond these structural changes, however, the solvent also functions as an energy source and sink, driving and relaxing nonadiabatic transitions---a dynamical feature which is not often fully explored in excited state nonadiabatic simulations. Of course, solvation and solvent dynamics are very complicated, and one can ask many different questions about such effects: how many molecules drive the downwards hop? How many molecules trap the energy? Are some solvent atoms more active than others at driving relaxation? How long does it take for electronic energy to be thermalized? In a future publication, we will analyze the role of solvent at promoting relaxation in a more detailed fashion.

For the moment, within the surface hopping protocol, we note that according to FSSH,  an electronic transition between states $i$ and $j$ is promoted by the $\vec{d}_{ij}\cdot\dot{\vec{R}}$ term in \refeq{eq:cit}. Thus, of the many questions listed above, the simplest question one can ask is: how delocalized are the $\vec{d}_{ij}\cdot\dot{\vec{R}}$ matrix elements? How many molecules actually drive the electronic transition downwards for CHD?

This question can be partially answered in the framework of a participation number. For a normalized distribution $\hat{w}=(\hat{w}_1, \hat{w}_2,\ldots, \hat{w}_N)$ ($\sum_k^N w_k = 1$), the participation number is defined as follows:
\begin{equation}
    p_n = \frac{1}{\sum_k^N \hat{w}_k^2}
\end{equation}
and gives an indication of how many components of the distribution contribute to the whole.
Consider the case of N equal weights, $w_i=1/N$, then $p_n=N$; and contrast with the case where $w_1=1$ and all other $w_i=0$, then $p_n = 1$.
The participation number, $p_n$, may be familiar by way of its relation to the inverse participation ratio, $i.p.r = p_n/N$, a measure to quantify localization of a wavefunction on a disordered lattice\cite{weaire1976:ipr}. In that context the $\hat{w}_k=c_k^2$, the expansion coefficients of the wavefunction, $\ket{\Psi}=\sum_k c_k \ket{k}$.

With this metric in mind, let us consider the quantity $\vec{d} \cdot \dot{\vec{R}}$ for each \emph{atom} in the system during a hopping event. Specifically, we compute
\begin{align}
    w_k &= \sqrt{\sum_{\gamma} {\qty(d^{k \gamma} \cdot \dot{R}^{k \gamma})}^{2}} \\
    \hat{w}_k &= \frac{w_k}{\sum_k^N w_k}
\end{align}
where $\gamma \in \{x,y,z\}$ indexes the Cartesian coordinates and $k$ indexes the atoms in the system.
in \fig{fig:gappartsuccess}, we plot both the energy gap and the \emph{participation number} of $\vec{d}_{ij}\cdot\dot{\vec{R}}$ at the time of a successful hop.
Given the lack of a solvatochromatic shift in \fig{fig:spectrum} above, one is perhaps not surprised that the energy gap distribution is largely similar for successful hops in \fig{fig:gappartsuccess}~(a): in all cases, the great majority of hops occur for small energy gaps ($<1~eV$). 
More interestingly, however, in \fig{fig:gappartsuccess}~(b), we find that distribution of the participation ratio is broader and shifted larger in solvent.
In other words, the solvent {\em is} clearly driving the transition downwards. Nevertheless, it appears that the solvent is not playing a crucial role; after all, the standard deviation of the participation numbers for successful hops are are 1.3 (vacuum), 1.4  (toluene), and 2.8 (ethanol).

\begin{figure}
    \centering
    \includegraphics[width=0.32\textwidth]{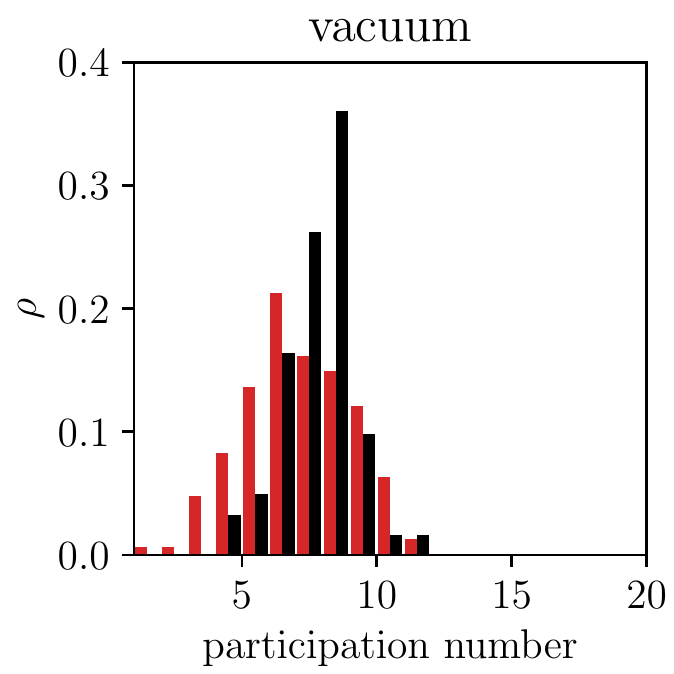}
    \includegraphics[width=0.32\textwidth]{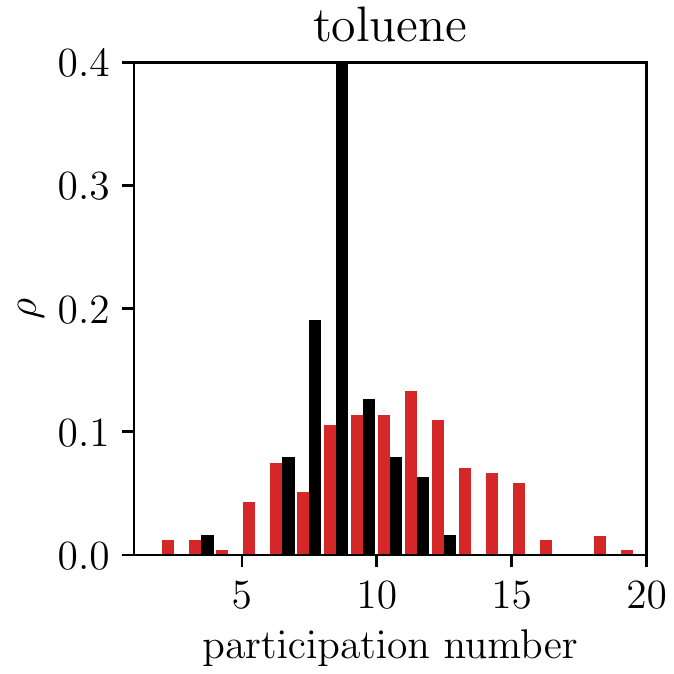}
    \includegraphics[width=0.32\textwidth]{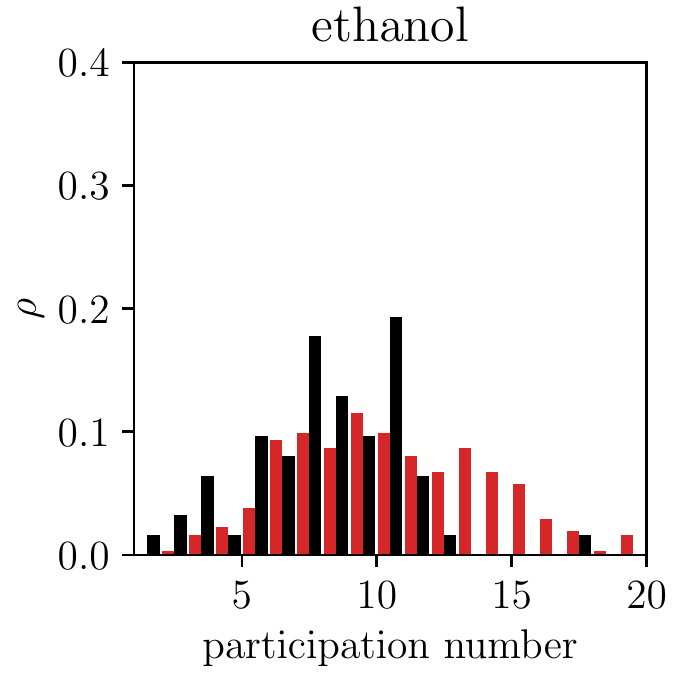}
    \caption{Distribution of successful (black) and failed (red) hops by the participation number of the derivative coupling between the $S_1$ and $S_0$ state for the three simulations in vacuum, in toluene and in ethanol. Distributions for successful hops are largely similar with means $\pm$ standard deviation:  $7.9  \pm 1.3$ (vacuum), $8.6  \pm 1.4$ (toluene), and $8.0  \pm 2.8$ (ethanol). Distributions for frustrated hops are substantially different for the solvated systems: $7.1  \pm  2.0$ (vacuum), $11.1 \pm  4.0$ (toluene), $11.3 \pm  4.8$ (ethanol).}

    \label{fig:partall}
\end{figure}

Therefore, for now (and within the limitations of a spin-flip electronic structure calculation), we tentatively conclude that the solvent is not driving the photochemical transition of CHD.  Instead, our current hypothesis is that the 10\% difference in HT yield between the vacuum and solvated environments (highlighted in \fig{fig:populations}) are the result of differences in vibrational energy dissipation after the hop downwards. Such a hypothesis has been motivated by investigating the nature of  \emph{frustrated} hops in our simulations.  Recall that frustrated hops are the essential ingredient that allows surface hopping to reach thermal equilibrium.

In \fig{fig:partall}, we compare and contrast the participation number distribution for successful downward hops versus those for frustrated hops.
Now, we see a dramatic broadening and shift to larger values in the distribution for frustrated hops when in solvent.
In other words, even though the solvent is not driving the electronic transition downwards, the solvent is attempting to drive an electronic transition upwards; but the delocalization of vibrational energy within the solvent forbid such a transition (in accordance with the second law of thermodynamics). 
In a future publication, we will explore in greater detail the nature of how the electronic energy is converted into vibrational energy and then delocalized across the solvent.

\section{Conclusion}\label{sec:conclusion}
We have presented INAQS, an Interface for Non-Adiabatic Quantum mechanics/molecular mechanics in Solvent. Among its demonstrated capabilities are single surface dynamics for the calculation of spectra, ground-state umbrella sampling in a QM/MM framework, and non-adiabatic surface hopping dynamics for studying electronic relaxation processes when coupled to a large environment. Here, we have studied the CHD molecule, but the most important applications in the future will no doubt investigate processes with large dipole moments and/or strong system-solvent interactions, especially charge transfer processes.

\begin{acknowledgement}
The authors thank Gerrit Groenhof and Dmitri Morozov for their help in understanding the Gromacs code.
MM and SF thank the Innovational Research Incentives Scheme Vidi 2017 with project number 016.Vidi.189.044, which is (partly) financed by the Dutch Research Council (NWO), for their funding.
This work was supported by the U.S. Air Force Office of Scientific Research (USAFOSR) under Grant Nos. FA9550-18-1-0497 and FA9550-18-1-0420, and the National Science Foundation under Grant No. CHE-2102071.
The authors thank the US DoD High Performance Computing Modernization Program for computer time.
\end{acknowledgement}

\appendix

\section{Computational Details}
\label{sec:computation}
Solvent parameters and equilibrated slabs were taken from Caleman \latin{et al.} \cite{Caleman2012a}, generously made available at \url{http://virtualchemistry.org}.
The solvent environment is modled \emph{via} a droplet approach: no periodic boundary conditions and an infinite Coulomb cutoff, bound by a frozen shell of solvent molecules approximately 5~\AA\ thick, restrained by Gromacs's LINCS implementation\cite{gromacs45}.
For the purpose of equilibration, a classical force field for the CHD ground was derived using the Q-Force package \cite{Sami2021}, for which the required potentials, gradients, and Hessians were computed at the CAM-B3LYP/cc-pVDZ level of theory with Q-Chem.
Following the work of one of the authors\cite{edison2020:chd} the spin-flip variant of time-dependent density functional theory (SF-TDDFT) using the BHHLYP functional and a correlation consistent double zeta basis set (cc-pVDZ) were used for all ground- and excited-state simulations.
The solvent's influence on CHD is incorporated via an electrostatic embedding scheme as described in Section~\ref{sec:theory:qmmm}.

\paragraph{Ground state AIMD}
The system was prepared by a purely classical equilibration of 2~ns with 2~fs timestep under periodic boundary conditions in cubic boxes of lengths 45~\AA\ (ethanol) and 55~\AA\ (toluene).
During equilibration, the system was held at 298.15K using a modified Berendsen thermostat with time constant, $\tau_t=100$~fs, as implemented in Gromacs. \cite{gromacs45}
From the end of the classical trajectory, a spherical region was excised and the solvent molecules greater than 16~\AA\ (ethanol) or 20~\AA\ (toluene) away from the CHD were frozen.
The droplet contained 664 total and 187 unconstrained (ethanol) and 558 total and 157 unconstrained (toluene) solvent molecules (see \reffig{fig:CHDscheme}).
Within the droplet, the system was relaxed via a QM/MM equilibration of 2~ps with 0.5~fs timesteps.
During the QM/MM relaxation, the same thermostat was applied, but exclusively to solvent molecules.

\paragraph{Surface hopping}
In vacuum, ethanol, and toluene, 51 independent surface hopping trajectories were computed with classical time step 0.5~fs for 500~fs---approximately twice the time required for all trajectories to reach the ground state.
For solvated systems, independent structures were drawn at 10~ps intervals from classical simulations with periodic boundary conditions in a box of length 60~\AA.
Independent structures for the vacuum case were drawn at 200~fs intervals from a single ground state \emph{ab initio} trajectory.
Spherical droplets were constructed as before by freezing solvent molecules greater than 25~\AA\ from the CHD. Droplets contained $992\pm12$ total and $352\pm10$ unconstrained (toluene) and $1849\pm20$ total and $738\pm14$ unconstrained (ethanol) solvent molecules.
No thermostatting was applied to any part of the system after a 2~ps QM/MM ground state equilibration at 298.15K.
For all trajectories, initial electronic amplitudes were taken to be $\{c_{S_0} = 0; c_{S_1} = 1\}$ and integrated as described above in Section~\ref{sec:theory:dynamics}, with a time step of at most $\frac{0.5}{20}=0.025$~fs and possibly smaller---see Jain and coworkers' prescription for selecting the electronic integration time step\cite{jain2016:afssh}.
Since the spin-flip method produces states that are functionally triplets as well as singlets, singlet states for dynamics were selected on-the-fly by tracking the value of $S^2  = \ev{\hat{S}^2}{\Psi_I}$. \cite{minezawa:gordon:2009:spinflip,herbert:2015:spinflip}
After computing the first 8 spin-flip states, their $S^2$ values are compared; because of evident spin contamination, we accept as singlets all states with $S^2 < 1.2$.
Dynamics were preformed on the lowest two SF-TDDFT singlet excited states, which correspond to the ground, $S_0$, and first excited, $S_1$ states, respectively. 

\paragraph{Performance}
Optimizations to the Q-Chem interface made surface hopping calculations substantially faster than they would have been without.
Using 16 cores of a 2.80 GHz Intel Xeon 6242  processor, INAQS computed vacuum surface hopping trajectories for CHD at an average rate of 1~ps per 12 hours.
When solvated in ethanol or toluene, 1~ps required 17 hours on average for the same cluster.
These times are 2-3 times faster than what one finds for a naive job submission.

\section{A practical user input}\label{sec:usage}
INAQS is configured using a file with key-value pairs in the INI format, which specifies all options for the type of dynamics and the settings for the electronic structure package (Q-CHEM); see \fig{fig:inaqsconfig} for an example.
Simulations use standard Gromacs input files (top, gro, mdp), the last of which  must indicate QM/MM via the group scheme.
The velocity-Verlet integrator (option \texttt{md-vv}) may be used for any of the dynamics modes and leapfrog (option \texttt{md}) for any except surface hopping.
Periodic boundary calculations are not supported at present.
We expect experienced Gromacs users with a functional Q-Chem install will have little trouble setting up and running calculations.
Code and usage instructions are available at \url{https://inaqs.github.io}

\begin{figure}
\begin{verbatim}
  [inaqs]
  runtype = fssh

  [fssh]
  qmcode = qchem
  excited_states = 1
  active_state = 1
  min_state = 0
  dtc = 0.5

  [qchem]
  spin_flip = 1
  track_states = 1
  buffer_states = 6
  basis = cc-PVDZ
  exchange = BHHLYP
  scf_algorithm = DIIS_GDM
  nthreads = 16
\end{verbatim}
\caption{Sample input file, \texttt{inaqs\_config.ini}, to be read by INAQS. The options \texttt{basis}, \texttt{exchange}, and \texttt{scf\_algorithm} are passed directly to Q-Chem and support the same set of options as Q-Chem.}
\label{fig:inaqsconfig}
\end{figure}
\bibliography{bibliography.bib}

\begin{mcitethebibliography}{109}
\bibitem[Lin and Truhlar(2006)Lin, and Truhlar]{Lin2006}
Lin,~H.; Truhlar,~D.~G. QM/MM: what have we learned, where are we, and where do
  we go from here? \emph{Theor. Chem. Acc.} \textbf{2006}, \emph{117},
  185\relax
\mciteBstWouldAddEndPuncttrue
\mciteSetBstMidEndSepPunct{\mcitedefaultmidpunct}
{\mcitedefaultendpunct}{\mcitedefaultseppunct}\relax
\EndOfBibitem
\end{mcitethebibliography}

\end{document}